# Nanoporous monolithic microsphere arrays have anti-adhesive properties independent of humidity

Anna Eichler-Volf[1], Longjian Xue[2,*], Alexander Kovalev[3], Elena V. Gorb[3], Stanislav N. Gorb[3,*] and Martin Steinhart[3,*]

[1] Institut für Chemie neuer Materialien, Universität Osnabrück, Barbarastr. 7, 49069 Osnabrück, Germany
[2] School of Power and Mechanical Engineering; Wuhan University; Donghu South Road 8, Wuhan, Wuchang, 430072; Hubei, China
[3] Functional Morphology and Biomechanics, Zoological Institute, Kiel University, Am Botanischen Garten 9, 24118 Kiel, Germany
* Correspondence: longjian.xue@uos.de; sgorb@zoologie.uni-kiel.de; martin.steinhart@uos.de

**Abstract:** Bioinspired artificial surfaces with tailored adhesive properties have attracted significant interest. While fibrillar adhesive pads mimicking gecko feet are optimized for strong reversible adhesion, monolithic microsphere arrays mimicking the slippery zone of the pitchers of carnivorous plants of the genus Nepenthes show anti-adhesive properties even against tacky counterpart surfaces. In contrast to the influence of topography, the influence of relative humidity (RH) on adhesion has been widely neglected. Some previous works deal with the influence of RH on the adhesive performance of fibrillar adhesive pads. Commonly, humidity-induced softening of the fibrils enhances adhesion. However, little is known on the influence of RH on solid anti-adhesive surfaces. We prepared polymeric nanoporous monolithic microsphere arrays (NMMAs) with microsphere diameters of a few 10 μm to test their anti-adhesive properties at RHs of 2 % and 90 %. Despite the presence of continuous nanopore systems through which the inner nanopore walls were accessible to humid air, the topography-induced anti-adhesive properties of NMMAs on tacky counterpart surfaces were retained even at RH = 90 %. This RH-independent robustness of the anti-adhesive properties of NMMAs significantly contrasts the adhesion enhancement by humidity-induced softening on nanoporous fibrillar adhesive pads made of the same material.

**PACS:** J0101

## 1. Introduction

Surfaces with generic anti-adhesive topographies have recently attracted significant interest. Low adhesion to counterpart surfaces, i.e. small pull-off forces (or adhesion forces) $F_{Ad}$, can be achieved by reduction of the actual contact area [1]. Reduction of the actual contact area can in turn be achieved by functionalization of sample surfaces with appropriate surface topographies. Topography-induced anti-adhesive behaviour can then be defined as the reduction of $F_{Ad}$ on a topographically patterned sample surface as compared to a flat reference surface. In general, roughness strongly reduces the real contact area between rigid counterpart surfaces because contact is formed only at a few protrusions. Hence, anti-adhesive behaviour is inherent to contacting rigid surfaces as long as they are not flat on an atomic scale. However, even rigid counterpart surfaces may form strong adhesive contact if they are flat enough, as exploited for techniques such as direct wafer bonding [2]. Sticky and compliant counterpart surfaces can adapt to rough topographies of sample surfaces characterized by feature sizes below a few microns, resulting in conformal contact formation. Thus, the real contact area may be larger than the macroscopic contour area [3], and adhesion may even be enhanced. Only if the feature size of the roughness of the sample surface is increased to a few 10 μm, elastic counterpart surfaces lose their ability to form conformal contact to the sample surface. Then, the real contact area is reduced, and the sample surface is anti-adhesive even towards sticky and compliant counterpart surfaces [4]. Artificial surfaces typically exhibiting a combination of self-cleaning behavior, superhydrophobicity and anti-adhesive properties have predominantly been designed by mimicking hierarchically structured plant surfaces, such as the surface of the lotus leaf [5-11]. Arrays of microspheres with radii $r_s$ of the order of a few 10 μm, such as monolithic arrays of solid polystyrene (PS) microspheres [4], are a suitable material platform to realize the first hierarchical structure level characterized by roughness with feature sizes of a few 10 μm. Various methods have been employed to introduce further hierarchical structure levels. Examples include plasma structuring [12], the transfer of non-contiguously packed microsphere arrays from Langmuir troughs to flat or topographically patterned substrates [13], spraying methods [14,15], spin coating [16], pressing microspheres onto gummed tape [17], generation of nanostructured layers on micropatterned substrates [18], two-step molding combined with initiated chemical vapor deposition [19] and swelling of polyamide in acetic acid [20].

We have recently investigated artificial, topographically patterned specimens consisting of the block copolymer polystyrene-*block*-poly(2-vinylpyridine) (PS-*b*-P2VP), which contained continuous, spongy nanopore systems with pore diameters of a few 10 nm [21-23]. While the majority component PS formed the internal structural scaffold, the aprotic but polar P2VP blocks were segregated to the PS-*b*-P2VP/air interface. Three surface topography types were studied: 1) macroscopically flat nanoporous monoliths; 2) nanoporous monolithic microsphere arrays (NMMAs) consisting of a mixture of nanoporous PS-*b*-P2VP microspheres with $r_s$ values of ~12.5 μm and ~22.5 μm (NMMA-12.5/22.5) [23] and 3) nanoporous fibrillar adhesive pads with fibrillar contact elements 300 nm in diameter [21,22]. The nanoporous fibrillar adhesive pads were designed for strong adhesion via the contact splitting principle [24,25]. The contact splitting principle underlies strong adhesive contact between a fibrillar adhesive pad and a rough counterpart surface and involves the formation of a large number of discrete focal contacts between the fibrillar contact elements and the rough counterpart surface. Since for detachment only a few focal contacts need to be loosened at the same time, contact splitting allows reversible formation of strong adhesive contact in a large number of successive attachment-detachment cycles. Whereas nanoporous fibrillar adhesive pads showed topography-induced adhesion enhancement [21], NMMAs-12.5/22.5 showed topography-induced anti-adhesive behaviour [23]. Moreover, the effect of liquids supplied through the continuous nanopore systems to contact interfaces strongly depends on the surface topography of the sample surfaces. In the case of a fibrillar surface topography, supply of mineral oil led to an increase in $F_{Ad}$ by one order of magnitude [22]. In contrast, on NMMA-12.5/22.5 supply of mineral oil resulted in a reduction of $F_{Ad}$ to 50 % of the value obtained under dry conditions [23].

The influence of humidity on the adhesive properties of topographically structured artificial surfaces has often been neglected. Buhl et al. studied the influence of humidity on the adhesive performance of flat poly(dimethyl siloxane) (PDMS) test surfaces and of arrays of flat-ended PDMS micropillars with diameters between 2.5 μm and 5 μm [26]. Humidity did not affect adhesion on flat PDMS reference samples. However, for arrays of thin pillars adhesion strongly decreased along with increasing humidity. On the other hand, humidity-induced softening of setal keratin reduces the stiffness of adhesive gecko foot hairs, which in turn results in increased adhesion [27]. In a previous work, we could show that $F_{Ad}$ on nanoporous fibrillar PS-*b*-P2VP adhesive pads with fibril diameters of 300 nm could be reversibly switched between a low-adhesion state at low relative humidity (*RH*) and a high-adhesion state at high *RH*; $F_{Ad}$ in the high-adhesion state was one order of magnitude higher than $F_{Ad}$ in the low adhesion state [21].

NMMAs are a promising platform for anti-adhesive surfaces that allow, in addition, supply of liquid to their contact surface. However, taking into account the significant differences in $F_{Ad}$ found for fibrillar adhesive pads at different *RH*s, the question arises to what extent the anti-adhesive properties of NMMAs depend on *RH*. We used NMMAs as a representative model of anti-adhesive microsphere arrays to study the dependence of adhesion on *RH*, because NMMAs consist of the same material and because NMMAs exhibit the same nanoscopic pore structure than the nanoporous fibrillar PS-*b*-P2VP adhesive pads investigated in previous studies [21,22]. It is obvious that the assessment of the adhesive performance of NMMAs in general must include the assessment of their adhesive performance at different *RH*s. Here we show that NMMAs consisting of microspheres with $r_s$ = 12.5 μm (NMMA-12.5), $r_s$ = 22.5 μm (NMMA-22.5) and NMMAs-12.5/22.5 show pronounced topography-induced anti-adhesive properties in a broad RH range between *RH* = 2 % and *RH* = 90 %. This outcome contrasts the results obtained on nanoporous fibrillar adhesive pads of the same material where $F_{Ad}$ strongly increased along with *RH*.

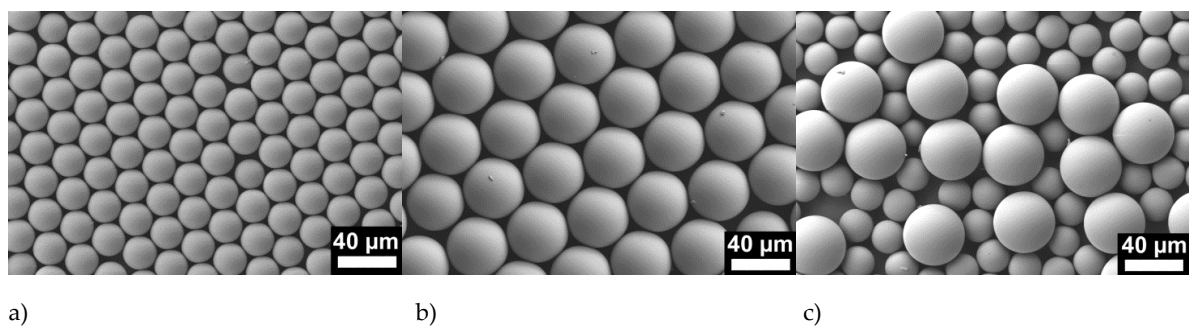

a)            b)            c)

**Figure 1.** Scanning electron microscopy images of monolayers of PS microspheres spin-coated on Si wafers, which were used as primary molds to prepare PDMS secondary molds. (a) $r_s$ = 12.5 μm; (b) $r_s$ = 22.5 μm; (c) $r_s$ = 12.5 and $r_s$ = 22.5 μm.

## 2. Results

### 2.1. Preparation of NMMAs

NMMAs were prepared by a double replication process. Following a procedure reported in reference [4], we prepared elastic secondary molds consisting of cross-linked PDMS. At first, PS monolayers used as primary molds were obtained by spin coating of PS microsphere suspensions on silicon (Si) wafer pieces with edge lengths of 1 cm. We prepared monolayers of PS microspheres with $r_s$ = 12.5 μm (Figure 1a) and with $r_s$ = 22.5 μm (Figure 1b) as well as mixed monolayers consisting of PS microspheres with $r_s$ values of 12.5 μm and 22.5 μm (Figure 1c). In the next step, asymmetric PS-*b*-P2VP with PS as major component and P2VP as minor component dissolved in tetrahydrofuran (THF) was deposited into the PDMS molds. After evaporation of the THF and nondestructive detachment from the PDMS molds, solid monolithic arrays of PS-*b*-P2VP microspheres connected to ~200 μm thick substrates of the same material were obtained. The solid monolithic PS-*b*-P2VP microsphere arrays were faithful positive replicas of the PS microsphere monolayers used as primary molds and likewise faithful negative replicas of the PDMS secondary molds. Thus, we obtained solid monolithic PS-*b*-P2VP microsphere arrays with $r_s$ = 12.5 μm (Figure 2a), with $r_s$ = 22.5 μm (Figure 2c) as well as mixed solid monolithic PS-*b*-P2VP microsphere arrays with $r_s$ values of 12.5 μm and 22.5 μm (Figure 2e).

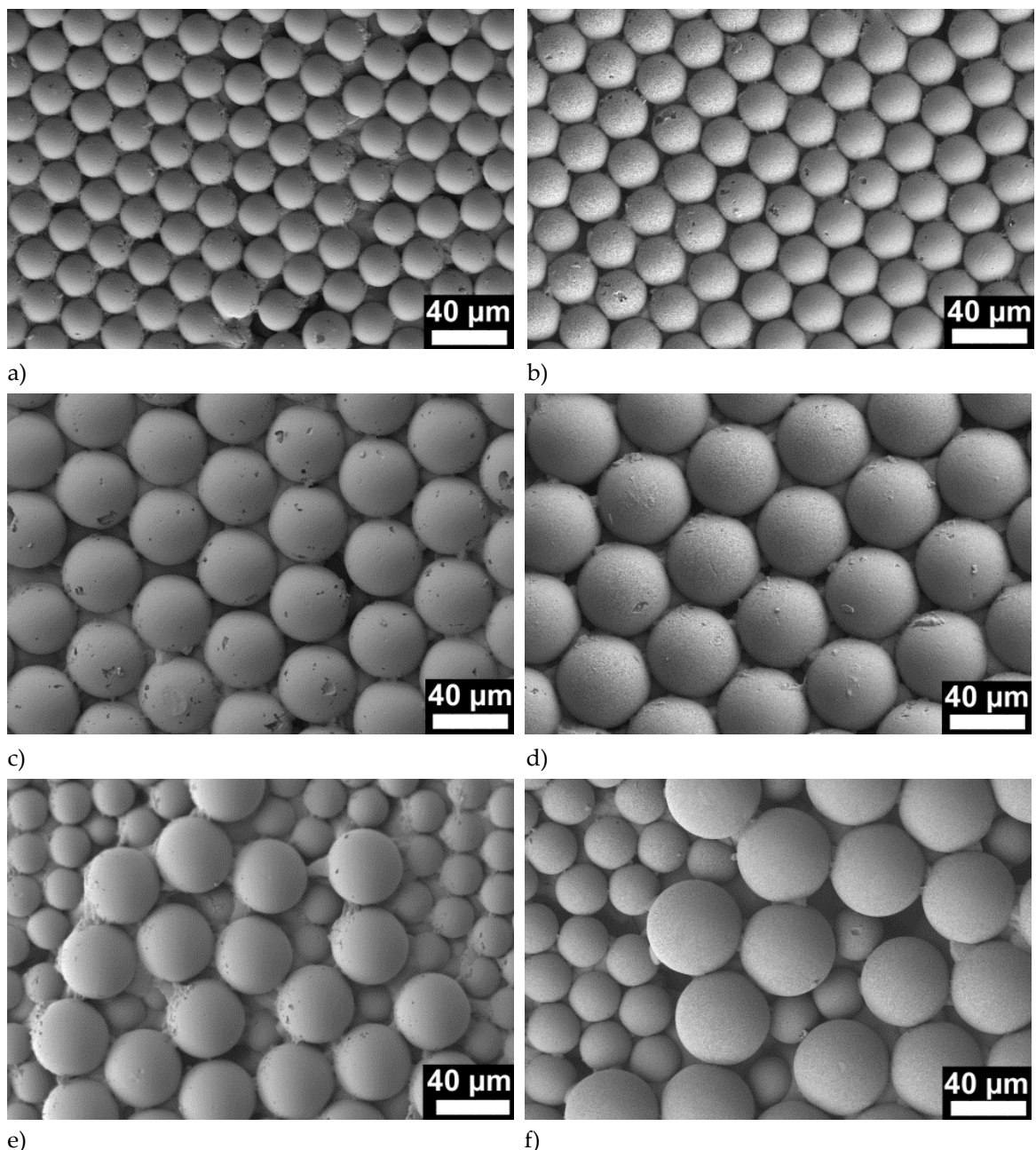

**Figure 2.** Scanning electron microscopy images of (a), (c), (e) solid monolithic arrays of PS-*b*-P2VP microspheres and of (b), (d), (f) NMMAs. The solid monolithic arrays of PS-*b*-P2VP microspheres shown in panels (a), (c) and (e) are positive replicas of the PS microsphere arrays displayed in Figure 1 that were used as primary molds. (b) NMMA-12.5 obtained from the solid monolithic array of PS-*b*-P2VP microspheres with $r_s$ = 12.5 µm displayed in panel (a). (d) NMMA-22.5 obtained by swelling-induced pore generation from the solid monolithic array of PS-*b*-P2VP microspheres with $r_s$ = 22.5 µm displayed in panel (c). (f) NMMA-12.5/22.5 obtained by swelling-induced pore generation from the solid monolithic array of mixed PS-*b*-P2VP microspheres with $r_s$ values of 12.5 µm and 22.5 µm displayed in panel (e). All images have the same magnification.

In the last preparation step, the solid monolithic PS-*b*-P2VP microsphere arrays were converted into NMMAs by swelling-induced pore generation [28-32] involving treatment with ethanol at 60 °C for 4 h. In brief, at 60 °C ethanol is a good solvent for P2VP but a non-solvent for PS. Osmotic pressure drives the ethanol molecules into the P2VP minority domains that consequently swell. The PS matrix in turn undergoes structural reconstruction processes to accommodate the increasing volume of the swelling P2VP domains. If the ethanol is allowed to evaporate, the extended P2VP blocks undergo entropic relaxation, whereas the glassy PS majority domains retain their reconstructed morphology. Swelling-induced pore generation at a given temperature can thus be stopped when the desired pore morphology has been reached. The pore size increases with increasing swelling time [29]. The kinetics of pore formation is also strongly influenced by temperature. At lower temperature, pore formation is slowed down, at higher temperature accelerated [30]. Under the conditions applied here, the volume increase results in the formation of a continuous network of the swollen P2VP domains. As a result, a continuous system of nanopores with walls consisting of relaxed P2VP chains is generated in place of the swollen P2VP domains (Figure 3). Porosimetry on NMMA12.5/22.5 revealed an average nanopore diameter of ~40 nm, a total nanopore volume of 0.05 cm$^3$/g and a specific surface area of 10 m$^2$/g [23]. These results can be regarded as representative for all samples investigated in this work. NMMAs derived from PS microsphere monolayers with $r_s$ = 12.5 μm are thereafter referred to as NMMA-12.5; NMMAs derived from PS microsphere monolayers with $r_s$ = 22.5 μm are thereafter referred to as NMMA-22.5; NMMAs derived from the mixed PS microsphere monolayers are thereafter referred to as NMMA-12.5/22.5. However, as obvious from comparisons of panels (a) and (b), of panels (c) and (d) as well as of panels (e) and (f) in Figure 2, swelling-induced pore generation increased the microsphere radii to ~120 % of the initial value. Flat nanoporous PS-*b*-P2VP monoliths were obtained by subjecting flat solid PS-*b*-P2VP films with a thickness of ~200 μm to the same swelling-induced pore generation procedure as the solid monolithic PS-*b*-P2VP microsphere arrays.

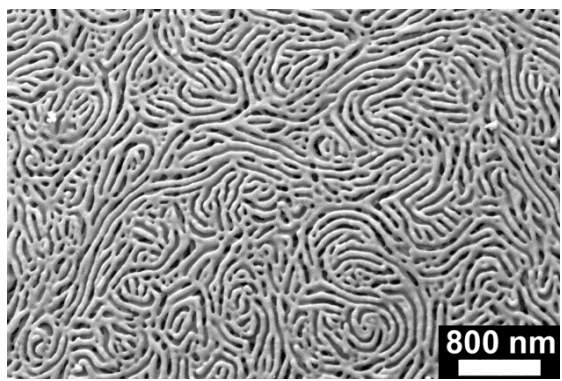

**Figure 3.** Representative scanning electron microscopy image of the surface of NMMA-12.5.

2.2. Pull-off force $F_{Ad}$

In this work, all force-displacement measurements were carried out in a humidity chamber at controlled *RH*s of either 2 % or 90 % (cf. Section 5.4 and Supporting Information to reference 21). We applied a technique specifically designed for the characterization of weakly adhesive surfaces [3]. A tacky PDMS half-sphere with a radius of 1.5 mm was displaced towards the sample surface to be tested until a loading force $F_L$ = 1.0 ± 0.1 mN was reached. Then, the PDMS half-sphere was immediately retracted. The retraction parts of representative force-displacement curves obtained on a flat nanoporous PS-*b*-P2VP monolith are shown in Figure 4. Because of the adhesion between the PDMS half-sphere and the tested sample surface, during retraction the PDMS half-sphere and the tested sample surface adhere to each other even beyond the displacement at which the applied force equals zero. While the PDMS half-sphere is retracted further, the pulling force reaches the value of the adhesion force or pull-off force $F_{Ad}$. In the retraction part of a force-displacement curve, $F_{Ad}$ corresponds to the global force minimum, as displayed in Figure 4 (note that per convention compressive forces have a positive sign and adhesive forces a negative sign). After the global force minimum corresponding to $F_{Ad}$ has been passed, the adhesive contact loosens and the adhesion between PDMS half-sphere and the tested sample surface disappears abruptly as the PDMS half-sphere is further retracted. Beyond the retraction length corresponding to $F_{Ad}$, the force immediately relaxes to the zero force line. This outcome indicates that there are, if at all, only negligible contributions of capillarity to adhesion. Similar force-displacement curves, albeit with smaller $F_{Ad}$ values (see below), were obtained on all NMMA species.

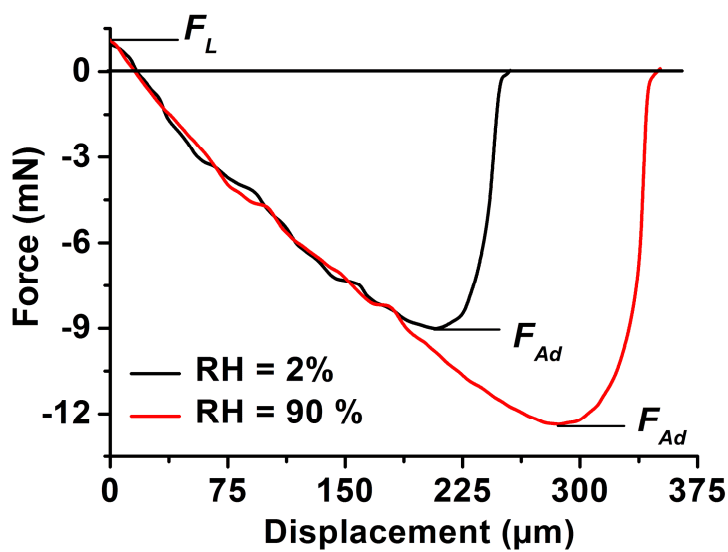

**Figure 4.** Retraction parts of force-displacement curves obtained on a flat nanoporous PS-*b*-P2VP monolith at *RH* = 2 % (black curve) and at *RH* = 90% (red curve). Retraction started immediately after a pre-set loading force $F_L$ of 1 mN ± 0.1 mN had been reached.

Figure 5 displays $F_L$ normalized to $F_L$ for flat nanoporous PS-*b*-P2VP monoliths and for NMMAs. At *RH* = 2 %, $F_{Ad}$ on NMMA-12.5 was reduced by a factor 2.3, on NMMA-22.5 by a factor of 2.4 and on NMMA-12.5/22.5 by a factor of 2.5 as compared to flat nanoporous PS-*b*-P2VP monoliths. $F_{Ad}$ on flat nanoporous PS-*b*-P2VP monoliths amounted to 9.3 ± 0.4 mN. On NMMA-12.5, $F_{Ad}$ amounted to 4.1 ± 0.2 mN, and on NMMA-22.5 to 3.9 ± 0.3 mN. $F_{Ad}$ on NMMA-12.5/22.5 was found to be 3.7 ± 0.5 mN. At *RH* = 90 %, $F_{Ad}$ on NMMA-12.5 was reduced by a factor 2.3, on NMMA-22.5 by a factor of 2.8, and on NMMA-12.5/22.5 by a factor of 4 as compared to flat nanoporous PS-*b*-P2VP monoliths. $F_{Ad}$ on flat nanoporous PS-*b*-P2VP monoliths amounted to 12.1 ± 0.4 mN. On NMMA-12.5, $F_{Ad}$ amounted to 5.2 ± 0.3 mN, and on NMMA-22.5 to 4.3 ± 0.3 mN. $F_{Ad}$ on NMMA-12.5/22.5 was found to be as low as 3.1 ± 0.7 mN.

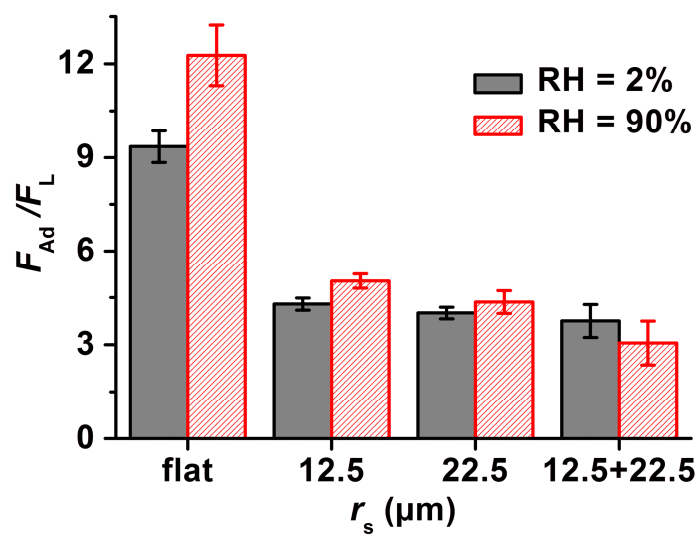

**Figure 5:** Average pull-off force $F_{Ad}$ normalized to the loading force $F_L$ = 1.0 mN ± 0.1 mN obtained on flat nanoporous PS-*b*-P2VP monoliths, NMMA-12.5, NMMA-22.5 and NMMA-12.5/22.5 at *RH* = 2 % and *RH* = 90%. The average $F_{Ad}/F_L$ values were obtained from at least 7 independent force-displacement measurements per sample and *RH* value. The error bars denote the standard deviations.

2.3 Work of separation $W_{Se}$

Besides $F_{Ad}$, the work of separation $W_{Se}$ can be extracted from the retraction parts of the force-distance curves. $W_{Se}$ is the area between zero force line and the negative-force part of the force-distance curves (Figure 4). $W_{Se}$ is the absolute amount of work required to detach the PDMS half-sphere from the surface of the tested sample. At RH = 2 %, $W_{Se}$ was found to be 1.40 ± 0.16 μJ for flat nanoporous PS-*b*-P2VP monoliths, 0.35 ± 0.03 μJ for NMMA-12.5 μm, 0.30 ± 0.04 μJ for NMMA-22.5 μm and 0.29 ± 0.05 μJ for NMMA-12.5/22.5 (Figure 6). At *RH* = 90%, $W_{Se}$ on a flat nanoporous PS-*b*-P2VP monoliths amounted to 2.58 ± 0.15 μJ, for NMMA-12.5 to 0.51± 0.06 μJ, for NMMA-22.5 to 0.42 ± 0.07 μJ and for NMMA-12.5/22.5 to 0.25 ± 0.08 μJ.

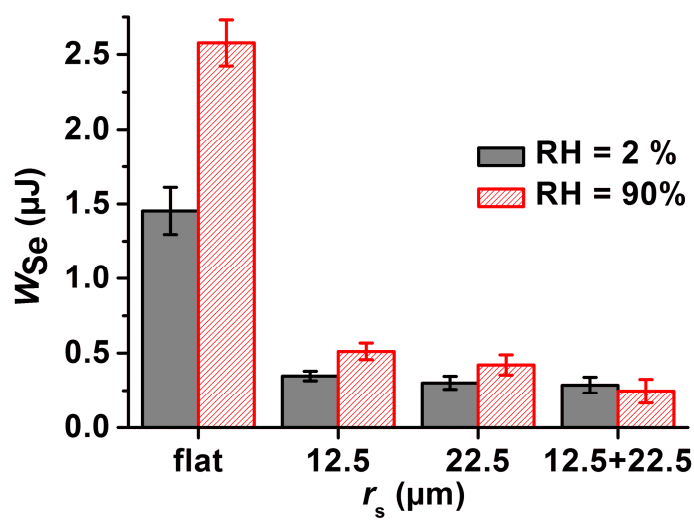

**Figure 6:** Average work of separation $W_{Se}$ obtained on flat nanoporous PS-*b*-P2VP monoliths, NMMA-12.5, NMMA-22.5 and NMMA-12.5/22.5 for a loading force $F_L$ = 1.0 mN ± 0.1 mN at *RH* = 2 % and *RH* = 90 %. The average $W_{Se}$ values were obtained from least 7 independent force-displacement measurements per sample and *RH* value. The error bars denote the standard deviations.

**3. Discussion**

It is obvious that the potentially pronounced dependence of adhesion on *RH* cannot be neglected when the performance and usability of functional surfaces is assessed. In general, one would expect that an increase in *RH* causes humidity-induced softening of most surfaces. According to the Johnson-Kendall-Roberts (JKR) theory [33], adhesion for a sphere-on-flat contact does not depend on the Young moduli of the materials in contact. However, on rough surfaces adhesion may increase with decreasing Young moduli of the materials in contact since the real contact area increases. Therefore, humidity-induced softening may result in enhanced adhesion. Force-displacement measurements with rigid spherical sapphire probes suitable for the measurement of strong adhesion on nanoporous fibrillar PS-*b*-P2VP adhesive pads revealed an increase in $F_{Ad}/F_L$ by nearly one order of magnitude if *RH* was increased from 2 % to 90 %. Moreover, on flat nanoporous PS-*b*-P2VP monoliths an increase in *RH* from 25 % to 90 % resulted in an increase in $F_{Ad}/F_L$ from 0.00 to ~0.09 for $F_L$ ~ 300 μN [21]. These results were explained by smaller effective elastic moduli of the PS-*b*-P2VP at high *RH* that caused adhesion enhancement. The humidity-induced increase in adhesion was leveraged by the presence of the internal nanopore structures. Apparently, the swelling-induced softening of the polar P2VP blocks was facilitated by the location of the P2VP blocks at the internal nanopore surfaces.

The surface topography of microsphere arrays with $r_s$ values of a few 10 μm is generically anti-adhesive for soft tacky PDMS [4]. Using tacky and compliant PDMS-half-spheres as probes, at *RH* = 42.6 % topography-induced reduction of $F_{Ad}/F_L$ was found on NMMA-12.5/22.5 [23]. In this work, we also used soft PDMS half-spheres as probes and found that anti-adhesive behaviour at *RH*s of 2 % and 90 % was evident not only on NMMA-12.5/22.5, but also on NMMA-12.5 and NMMA-22.5 (Figure 5). However, adhesion on flat nanoporous PS-*b*-P2VP monoliths increases as *RH* increases; independent of whether rigid spherical sapphire probes [21] or tacky PDMS half-spheres were employed, on flat nanoporous PS-*b*-P2VP monoliths increases in *RH* caused increases in $F_{Ad}/F_L$. The changes in $F_{Ad}/F_L$ on the NMMAs related to increases in *RH* from 2 % to 90 % were much less pronounced than on flat nanoporous PS-*b*-P2VP monoliths. In the cases of NMMA-22.5 and NMMA 12.5/22.5 this change was even not significant (t-test: $P > 0.05$). Even at *RH* = 90 %, all NMMAs showed anti-adhesive properties as compared to flat nanoporous PS-*b*-P2VP specimens at *RH* = 2 % (Figure 5). Similar trends were apparent for $W_{Se}$ (Figure 6). While a strong increase in $W_{Se}$ occurs on flat nanoporous PS-*b*-P2VP monoliths if *RH* is increased from 2 % to 90 %, an increase in *RH* from 2 % to 90 % causes only weak changes in $W_{Se}$ on NMMA-12.5 and NMMA-22.5. On NMMA-12.5/22.5 the mean values of $F_{Ad}/F_L$ and $W_{Se}$ even decrease if *RH* is increased from 2 % to 90 %.

The question arises whether capillarity effects influence adhesion at high *RH*s. Capillarity contributions would be apparent from characteristic tails at retraction lengths beyond the retraction length corresponding to the force minimum $F_{Ad}$ (cf. Figure 6 in reference [22]). These tails represents the work required to stretch and rupture liquid bridges between the separating surfaces. As apparent from the retraction parts of the force-displacement curves measured with tacky PDMS half spheres (Figure 4), significant capillarity contributions were absent even at *RH* = 90 %. The absence of pronounced capillarity effects was found for flat nanoporous PS-*b*-P2VP monoliths as well as for all NMMA species studied here. Beyond the displacement corresponding to the force minimum $F_{Ad}$, the force-distance curves immediately relax to the zero-force line. This outcome is in line with the results obtained with rigid spherical sapphire probes on flat nanoporous PS-*b*-P2VP monoliths as well as on nanoporous fibrillar PS-*b*-P2VP adhesive pads [21].

We obtained apparent elastic moduli of 105.7 ± 6.2 kPa (*RH* = 2 %) and 74. 2 ± 2.7 kPa (*RH* = 90 %) for the system "flat nanoporous PS-*b*-P2VP monolith"/"PDMS half sphere". While previously conducted force-displacement measurements using rigid spherical sapphire probes unambiguously revealed that nanoporous PS-*b*-P2VP softens if RH is increased [21], we cannot exclude that the PDMS half-spheres used as probes in this work may soften also if *RH* is increased. However, this softening would additionally increase adhesion at high *RH* values. Hence, the resilience of the NMMAs against humidity-induced increases in adhesion would even be underestimated. The question arises as to the origins of this resilience. The PDMS half-spheres used as probes form only focal contacts to the NMMAs at the caps of the microspheres. If the PS-*b*-P2VP forming the NMMAs gets softer at high *RH* and the microsphere caps form contact to the PDMS half-sphere, the microsphere caps might deform to some extent. However, this means that the PDMS half-sphere deforms less on contact formation so that the overall contact area does not significantly increase if RH is increased.

## 4. Conclusion

In this work, polymeric nanoporous monolithic microsphere arrays with microsphere diameters of a few 10 μm were prepared to test their anti-adhesive properties as function of relative humidity. The nanoporous monolithic microsphere arrays showed robust, topography-induced anti-adhesive properties on sticky counterpart surfaces independent of relative humidity. This result is in contrast to the pronounced dependence of adhesion on humidity found on flat nanoporous monoliths and on nanoporous fibrillar adhesive pads [21] made of the same material. Hence, the impact of relative humidity on adhesion is different for different surface topographies. The conservation of the anti-adhesive behavior toward sticky counterpart surfaces even at relative humidities of 90 % suggests that polymeric monolithic microsphere arrays can be used as robust and generic anti-adhesive surfaces under a broad range of operational conditions. Since this is true even for nanoporous monolithic microsphere arrays that allow supply of liquid to the contact interface through the nanopore system, further tailoring of adhesion through switching between dry and wet adhesion is possible over a wide range of relative humidities.

## 5. Materials and Methods

### 5.1 Sample preparation

PDMS primary molds were prepared following the procedures described in reference [4]. In brief, aqueous suspensions of PS microspheres (Polybead Microspheres, Polysciences Inc., Canada) with $r_s$ = 12.5 μm (2.91 x $10^6$ microspheres/ml) and $r_s$ = 22.5 μm (4.99 x $10^5$ microspheres/ml) were deposited onto rectangular pieces of (100) Si wafers with edge lengths of ~1 cm by applying a multistep spin coating program. Monolayers containing a mixture of PS microspheres with $r_s$ = 12.5 μm and $r_s$ = 22.5 μm were obtained by alternating spin-coating of the corresponding PS microsphere suspensions onto the Si wafer pieces. The obtained PS microsphere monolayers (Figure 1) were covered by a ~0.5 cm thick layer of PDMS prepolymer mixture (Sylgard 184, Dow Corning), which was prepared by mixing base and curing agent at a weight ratio of 10:1. After curing for 3 days under ambient conditions, the Si wafer pieces were mechanically detached; residual PS microspheres were removed mechanically and by sonication. Asymmetric PS-*b*-P2VP ($M_n$ (PS) = 101 000 g/mol; $M_n$(P2VP) = 29 000 g/mol; $M_w/M_n$ = 1.60; volume fraction of P2VP 21 %; bulk period ~51 nm) was obtained from Polymer Source Inc., Canada. Solutions containing 200 mg PS-*b*-P2VP per 2 mL tetrahydrofurane (THF) were deposited onto the PDMS molds and the THV was allowed to slowly evaporate. Detachment from the PDMS molds yielded monolithic, solid PS-*b*-P2VP microsphere arrays tightly connected to ~200 μm thick PS-*b*-P2VP substrates over areas of 5 mm x 5 mm (Figure 2a, c, e). Swelling-induced pore generation with ethanol at 60°C for 4 h yielded NMMAs (Figure 2b, d, f and Figure 3). Flat nanoporous PS-*b*-P2VP monoliths with a thickness of ~200 μm were prepared as described in reference [23].

### 5.2 Scanning electron microscopy (SEM)

For the SEM investigations, we used a scanning electron microscope Zeiss Auriga operated at an accelerating voltage of 5 kV. Prior to the SEM investigations, the samples were sputter-coated with a thin layer of gold-palladium alloy.

### 5.3 Force-displacement measurements

All force-displacement measurements were carried out at a temperature of 23.1°C. The PDMS half-spheres with a radius of 1.5 mm used as probes in the force-displacement measurements were prepared following a procedure described elsewhere [3,23]. For each sample, a new PDMS half-sphere from a set of identical PDMS half-spheres was used. Force-displacement measurements were carried out using a home-made microforce biotester Basalt-01 and Basalt BT V 3.0 (+VISAM) software (TETRA GmbH, Ilmenau, Germany). The PDMS half-spheres were glued to a metal spring with a spring constant of 238.93 N/m. The PDMS half-spheres were displaced towards the tested sample surface until a preset loading force $F_L$ = 1±0.1 mN was reached and then immediately retracted. The approach and retraction speeds were 50 μm/s. The applied force as well as the displacement of the PDMS half-sphere was recorded. $W_{Se}$ was determined from the retraction parts of the force-displacement curves by applying the trapezoidal rule (Newton-Cotes formula) [34] using Matlab 7.10 (MathWorks, Natick, NA, USA). We calculated the apparent elastic moduli of the system "flat nanoporous PS-*b*-P2VP monolith"/"PDMS half sphere" by fitting the retraction parts of the force-displacement curves using equation 9 derived in reference [35] corresponding to the JKR theory [33]. For each of sample / *RH* combination at least seven force-displacement measurements were performed and subsequently analyzed.

5.4 Humidity control during force-displacement measurements

The force-displacement measurements were carried out in a closed homemade humidity chamber. The desired *RH* was adjusted by appropriate mixing ratios of dry and humid nitrogen gas, as described in detail and schematically displayed in the Supporting Information to reference [21]. A part of the dry nitrogen gas (*RH* = 2 %) was directly piped into gas washing bottle 2. Another part of the dry nitrogen gas was piped through gas washing bottle 1, which was filled with water. Thus, humid nitrogen gas (*RH* = 90 %) was obtained, which was then piped from gas washing bottle 1 into gas washing bottle 2. A system of valves allowed adjusting the flow rates of dry and humid nitrogen gas into gas washing bottle 2. In this way, the mixing ratio of dry and humid nitrogen gas and, consequently, the desired *RH* could be adjusted in gas washing bottle 2. The nitrogen gas with the desired *RH* was subsequently piped into gas washing bottle 3 where *RH* was measured by means of a hygrothermograph. Finally, the nitrogen gas with the desired *RH* was piped from gas washing bottle 3 into the humidity chamber where the forced-displacement measurements were performed. The volume of the gas washing bottles 1-3 amounted to 1 L. Prior to the force-displacement measurements, *RH* in the humidity chamber was equilibrated for 15 minutes.

**Acknowledgments:** Support by the German Research Foundation (DFG Priority Program 1420 "Biomimetic Materials Research: Functionality by Hierarchical Structuring of Materials") and by the European Research Council (ERC-CoG-2014; project 646742 INCANA) is gratefully acknowledged.

**Author Contributions:** M.S., L.X. and S.N.G conceived and designed the experiments; A. E.-V. performed the experiments; A.E.-V., A.K. and E.V.G. analyzed the data; M.S. wrote the paper; all authors approved the submitted version of the paper.

**Conflicts of Interest:** The authors declare no conflict of interest.

**Abbreviations**

The following abbreviations are used in this manuscript:

$F$: Force
$F_{Ad}$: adhesion force or pull-off force
$M_n$: number-average molecular weight
$M_w$: weight-average molecular weight
NMMA: nanoporous monolithic microsphere array
PDMS: poly(dimethyl siloxane)
PS: polystyrene
PS-*b*-P2VP: polystyrene-*block*-poly(2-vinyl pyridine)
P2VP: poly(2-vinyl pyridine)
*RH* = relative humidity
$r_s$: microsphere radius
Si: silicon
THF: tetrahydrofurane
$W_{Se}$: work of separation